\documentclass[11pt]{article}
\usepackage{epsfig}
\newcommand{\noi}{\noindent}

\def\be{\begin{equation}}
\def\ee{\end{equation}}
\def\bea{\begin{eqnarray}}
\def\eea{\end{eqnarray}}

\def\g5{\gamma_5}
\def\gm{\gamma_\mu}
\def\pa{\partial}

\def\ot{(1\,\leftrightarrow\,2)}
\def\sa{\vec \sigma_1}
\def\sb{\vec \sigma_2}

\def\qa{\vec q_1}
\def\qb{\vec q_2}
\def\q2{\vec q_2}
\def\vq{\vec q}
\def\Pa{\vec P_1}

\def\vamn{\varepsilon^{a m n}}
\def\s2q2{(\vec \sigma_2 \times \vec q_2)}

\def\tb{\,\tau^{\,a}_2}
\def\t1t2a{\,i\,(\vec \tau_1 \times \vec \tau_2)^a}
\def\tb3{\,\tau^{\,3}_2}
\def\t1t23{\,i\,(\vec \tau_1 \times \vec \tau_2)^3}
\def\fot{\frac{1}{2}}
\def\Journal#1#2#3#4{{#1} {\bf #2}, #3 (#4)}
\def\NPA{{\em Nucl. Phys.} A}
\def\NPB{{\em Nucl. Phys.} B}
\def\PRL{\em Phys. Rev. Lett.}
\def\PRC{{\em Phys. Rev.} C}

\def\PRW{\em Phys. Rev.}
\def\FBS{\em Few--Body Systems}
\def\PLB{{\em Phys. Lett.} B}
\def\PR{\em Phys. Rep.}
\def\IJMPA{{\em Int. J. Mod. Phys.} A}
\def\APNY{\em Ann. Phys. (N.Y.)}
\def\NC{\em Nuovo Cim.\,\,}
\def\ANP{\em Adv. Nucl. Phys.\,\,}
\def\JPG{{\em J. Phys.}  G}
\def\AJ{\em Astrophys. J.\,\,}
\def\PPNP{\em Prog. Part. Nucl. Phys.\,\,}
\def\HPHA{\em Helv. Phys. Acta\,\,}
\date{}
\begin{document}
\title{\bf The role of the pion pair term in the theory of the
           weak axial meson exchange currents }
\author{
B.~Mosconi \\
{\small Universit$\grave{a}$ di Firenze, Department of Physics,
and Istituto Nazionale di Fisica Nucleare, Sezione di Firenze,} \\
{\small I-50019, Sesto Fiorentino (Firenze), Italy,
        ~e-mail: mosconi@fi.infn.it}
\vspace{0.5 cm} \\
P.~Ricci \\
{\small Istituto Nazionale di Fisica Nucleare, Sezione di Firenze,}\\
{\small  I-50019, Sesto Fiorentino (Firenze), Italy, ~e-mail: ricci@fi.infn.it}
\vspace{0.5 cm} \\
E.~Truhl\'{\i}k \thanks{Talk presented
           at the workshop "Nuclear Forces and QCD: Never the
           Twain Shall Meet?", ECT* Trento: June 20th--July 1st
           2005.}\\
{\small Institute of Nuclear Physics, Academy of Sciences of the
Czech
Republic,} \\
{\small CZ--250 68 \v{R}e\v{z}, Czechia, ~e-mail: truhlik@ujf.cas.cz}
}
\maketitle
\begin{abstract}
\noi The structure of the weak axial pion exchange current is
discussed in various models.  It is shown how the interplay of the
chiral invariance and the double counting problem restricts
uniquely the form of the pion potential term, in the case when the
nuclear dynamics is described by the Schr\"odinger equation with
static nucleon-nucleon potential.

\end{abstract}
\bigskip\smallskip
\noi \hskip 1.cm {\bf PACS.} 11.40.Ha Partially conserved axial-vector
     currents   \\
\hspace*{2.4cm} 25.30.-c Lepton-induced nuclear reactions

\noi \hskip 1.cm {\bf Key words.} weak axial - nuclear - pion exchange current
\newpage
\section{Introduction}\label{intro}

The semileptonic weak nuclear interaction has been studied for
half a century. The basic cornerstones of this field of research
are (i) the chiral symmetry, (ii) the conserved vector current and
(iii) the partial conservation of the axial current (PCAC). In the
formulation \cite{SLA}, the PCAC reads
\be
q_\mu\,<\Psi_f|j^a_{5\mu}(q)|\Psi_i>\,=\,if_\pi
 m^2_\pi\Delta^\pi_F(q^2)\, <\Psi_f|m^a_\pi(q)|\Psi_i>\,,
 \label{PCAC}
\ee
where $j^a_{5\mu}(q)$ is the total weak axial
isovector current, $m^a_\pi(q)$ is the total pion source (the pion
production/absorption amplitude) and $|\Psi_{i,f}>$ is the wave
function describing the initial (i) or final (f) nuclear state.
\par
It has been recognized \cite{BS} in studying the triton beta decay
\be ^{3}H\,\rightarrow\,^{3}He\,+\,e^-\,+\,{\bar \nu}\,\,  \label{TBD}
\ee
that in addition to the one--nucleon current, the effect of the
space component of weak axial exchange current (WAEC) enhances the
Gamow--Teller matrix element that is to be compared to the one
extracted from the data. This suggests that the current
$j^a_{5\mu}(q)$ can be understood for the system of $A$ nucleons
as the sum of the one- and two--nucleon components,
\be
j^a_{5\mu}(q)\,=\,\sum^A_{i=1}\,j^{a}_{5\mu}(1,i,q_i)\,+\,\sum^A_{i<j}\,
j^{a}_{5\mu}(2,ij,q)\,.  \label{jtot}
\ee
\par
Let us describe the
nuclear system by the Schr\"odinger equation
\be
 H|\Psi>\,=\,E|\Psi>\,, \label{NEM}
\ee
with the Hamiltonian H,
\be
 H\,=\,T\,+\,V\,, \label{H}
\ee
where $T$ is the kinetic energy and
$V$ is the nuclear potential describing the interaction between
nucleon pairs. Taking for simplicity $A=2$, we obtain from
Eq.\,(\ref{PCAC}) in the operator form and from Eqs.\,(\ref{jtot})
and (\ref{H}) the following set of equations for the one- and
two--nucleon components of the total axial current
\bea \vec q_i
\cdot \vec j^a_{\,5}(1,\vec q_i)\,&=&\,[\,T_i\,,\,\rho^a_{\,5}
(1,\vec q_i)\,]\,+\,if_\pi m^2_\pi
\Delta^\pi_F(q^2)\,m^a_\pi(1,\vec q_i)\,,\quad i=1,2\,,  \label{NCEoi}  \\
\vec q \cdot \vec j^a_{\,5}(2,\vec q)\,&=&\,[\,T_1+T_2\,,\,
\rho^a_{\,5}(2,\vec q)\,]\,+\,([\,V\,,\,\rho^a_{\,5}(1,\vec q)\,]+\ot)
                                                          \nonumber \\
& &+\,if_\pi m^2_\pi \Delta^\pi_F(q^2)\,m^a_\pi(2,\vec q)\,.
\label{NCEt}
\eea
\par
In Eq.\,(\ref{NCEt}), we neglected $\rho^a_{\,5}(2,\vec q)$ in the
second commutator on the right hand side. If the WAEC is
constructed so that it satisfies Eq.\,({\ref{NCEt}), then the
matrix element of the total current, sandwiched between solutions
of the Schr\"odinger equation (\ref{NEM}), satisfies the PCAC
(\ref{PCAC}).
\par
It is known from the dimensional analysis \cite{KDR}, that the
space component of the WAEC  $\vec j^a_{\,5}(2,\vec q)$ is of the
order ${\cal O}(1/M^3)$ (M is the nucleon mass). Being of a
relativistic origin, it is model dependent. This component of the
WAEC was derived by several authors in various models. In the
standard nuclear physics approach the model systems of strongly
interacting particles contain various particles, such as baryons
$N$, $\Delta(1232)$, pions and heavy mesons
\cite{IT,ISTPR,AHHST,TK1,Sci,TR,SMA}. On the other hand, in
effective field theories, one uses Lagrangians with the heavy
particle degrees of freedom integrated out and preserving nucleon,
delta and pion \cite{HHK} or nucleon and pion \cite{PMR,PKMR} or
only nucleon \cite{KSW,MB2} degrees of freedom.
\par
Accepting the chiral symmetry as the basic symmetry governing the
nuclear dynamics, it is expected that the WAEC of the pion range,
constructed within approaches respecting this symmetry and in
conjunction with the given nuclear equation of motion, should
exhibit model independence. On the other hand, checking the weak
axial pion exchange currents, constructed in
\cite{ISTPR,AHHST,TK1,Sci,TR,SMA,PKMR,MB2} one concludes that the
situation is not transparent. Let us discuss various approaches in
more detail.
\par
Let us first classify, in general, the WAEC of a given range into
potential and non-potential currents. In analogy with the
electromagnetic sector, the potential WAEC is such that it
satisfies the part of Eq.\,(\ref{NCEt}) containing the commutator
\mbox{$[\,V\,,\,\rho^a_{\,5}(1,\vec q)\,]$}. As in the case of the
electromagnetic interaction, the pair term is one of the exchange
currents that belong to the potential current. It is obtained by
the non--relativistic reduction of the negative frequency part of
the nucleon Born term. Besides, other potential currents can
appear. Then the total potential WAEC is defined as the sum of all
potential terms of a given range.
\par
The approach of Ref.\,\cite{TR} in constructing the WAEC is the
only one that is not based on chiral Lagrangians. It uses the
relativistic nucleon Born terms and the WAEC is obtained by
imbedding the nuclear potential into the negative frequency part
of these terms, thus directly connecting the potentials and the
WAEC. It follows that this current is not chiral since it is not
based on any chiral group. The pion pair term is obtained using
the pseudovector (PV) $\pi NN$ coupling \footnote{Being of the
order ${\cal O}(1/M^5)$, it is negligible.}, which could be
considered as an argument that the global chiral invariance is
respected. However, any model is chiral invariant only if the
resulting current does not depend on the choice of the $\pi NN$
coupling, which is not the case of Ref.\,\cite{TR}. As discussed
in the last paragraph of Sect.\,2 of Ref.\,\cite{TR}, this
construction does not tolerate the pseudoscalar (PS) $\pi NN$
coupling, since it provides the weak pion production amplitude
that is at variance with the current algebra prediction. However,
it was overlooked in \cite{TR} that the chiral model \cite{IT}
with the PS $\pi NN$ coupling does provide the correct weak pion
production amplitude. In other words, one should use chiral models
and not simple $\pi NN$ couplings.
\par
In Ref.\,\cite{AHHST}, the WAEC is derived within the extended
S--matrix method \cite{ATA}, using the chiral Lagrangian model
with the PV $\pi NN$ coupling \cite{IT} . The resulting potential
current is of the order ${\cal O}(1/M^3)$ and is given by the
difference of the nucleon Born term and the first Born iteration.
In Ref.\,\cite{TK1}, the same potential current is obtained from
the chiral model \cite{IT} with the PS $\pi NN$ coupling. In this
case, besides the pair term, the PCAC constraint term contributes.
\par
The pion pair term of Refs.\,\cite{Sci,ELM} is derived from the
PS $\pi NN$ coupling that is not chiral invariant, as it was correctly
noted in Ref.\,\cite{ASP}.
\par
In Ref.\,\cite{PKMR}, the WAEC is derived within the heavy
baryon chiral perturbation theory (HBChPT) approach, but the pion
pair term is considered as fully reducible and therefore omitted.
\par
Moreover, it is not clear from Refs.\,\cite{Sci,TR,PKMR} that the
constructed WAEC of the pion range satisfies a particular form of
the PCAC in conjunction with a specific nuclear equation of
motion\footnote{According to Ref.\,\cite{ELM}, the axial current
is not supposed to satisfy any continuity equation, in contrast to
the electromagnetic current.}. In other words, the problem of
double counting is overlooked. As it will become clear later,
these currents do not satisfy the PCAC as stated in
Eq.\,(\ref{PCAC}), if used in standard nuclear physics
calculations, based on the Schr\"odinger equation and static
nuclear potentials.
\par
Here we shall discuss the role of the weak axial pion pair term in
fulfilling Eq.\,(\ref{NCEt}) in conjunction with the Schr\"odinger
equation and the static nuclear potential. Simultaneously, we
shall consider the problem of double counting. We shall construct
the pion pair term and the related potential current in two
models. In Sect.\,\ref{secII.1}, we start from the Lagrangian of
the $\pi N$ system used in the chiral perturbation theory, from
which we construct in the leading order (tree approximation) the
WAEC of the pion range. We explicitly show how the potential and
non--potential parts interplay with other components entering
Eq.\,(\ref{NCEt}) so that the continuity equation is satisfied.
The resulting potential term is the same as the one derived
earlier in \cite{IT,ISTPR,AHHST,TK1} from the hard pion Lagrangian
of the $N\Delta\pi\rho a_1$ system. In Sect.\,\ref{secII.2}, we
derive the potential term in the leading order of the HBChPT
approach. We show that the obtained current is the same as the one
derived in Sect.\,\ref{secII.1}. We compare the space component of
the long--range part of the WAEC computed in various models in
Sect.\,\ref{secII.3}, where we also calculate the effect of the
potential term in the  deuteron weak disintegration by low energy
neutrinos in the neutral current channel. In Sect.\,\ref{secIII},
we summarize our results.

\section{The pion pair term and the nuclear PCAC } \label{secII}

In constructing the weak axial potential pion exchange operator we
start from the set of relativistic Feynman amplitudes satisfying
the PCAC equation. Generally, these amplitudes are not yet the
nuclear exchange currents, because of the double counting
problem: the presence of the pair term in the exchange current
operator is related to the equation, describing the nuclear
states. If the nucleon propagator in the first Born iteration is
the full relativistic one then this iteration is equal to the
nucleon Born term and the exchange currents do not contain any
pair term, in order to avoid the double counting. This is the
case of the axial currents constructed  in conjunction with the
Bethe--Salpeter equation \cite{KT1}. In this case, the nucleon
Born term is fully reducible. In the case of the Schr\"odinger
equation, the nucleon Born term is not fully reducible. The
propagator of the first Born iteration contains only the positive
frequencies and usually, the  nuclear potential is the static one.
Then the negative frequency part is not the only one to contribute
to the exchange current operator from the nucleon Born term. If
the Feynman amplitudes are constructed using the chiral model with
the PV coupling, the positive frequency part of the nucleon Born
term does not coincide with the first Born iteration and  the
difference should be calculated. Then the resulting potential
current is equal to this difference, since the negative frequency
part of the nucleon Born term (the pair term) is suppressed by a
factor $\approx\,1/M^2$ and therefore, negligible.
\par
Let us note that the method of the construction of the nuclear
WAEC \cite{ISTPR,AHHST,TK1} we use here
 was considered earlier \cite{ATA} also for the construction
of the electromagnetic exchange currents. The resulting exchange
currents, $j_\mu(2,q)$, containing the leading relativistic
corrections in both the space and time components, satisfy the
current conservation constraint
\be
q_\mu j_\mu(2)\,=\,([\,V\,,\,\rho(1,\vec q)\,]+\ot)\,, \label{NCVC}
\ee
\par
These currents coincide with the exchange currents derived within
the framework of the transformation method as it was shown in
Ref.\,\cite{GA}.
\par
Below we shall use two model Lagrangians of the $\pi N$ system
that also include the external electroweak fields aiming to
demonstrate the appearance of the potential term of the same order
in $1/M$ as other axial pion--exchange currents, and to show that
its presence is required by the PCAC hypothesis. The first model
Lagrangian is the basic Lagrangian of the chiral perturbation
theory \cite{S}. The second model Lagrangian is that of the HBChPT
used in \cite{PMR} to construct the WAEC. We construct the
currents in the leading order only, since the higher order
corrections cannot change our conclusions.

\subsection{The weak axial pion pair term within the formalism of
the chiral perturbation theory} \label{secII.1}

We start from the Lagrangian of the $\pi N$ system
\cite{S,BKY,STG}
\be
{\cal L}_{\pi N}\,=\,-{\bar N}\,\gm\,(\pa_\mu-i{\bar
\alpha}_{\mu\,\|})\,N\,-\,M\,{\bar N}N\,+\,ig_A{\bar N}\,
\gm\g5\,{\bar \alpha}_{\mu\,\perp}\,N\,,  \label{LPNNL}
\ee
\be
\bar \alpha_\mu\,=\,-i[\pa_\mu\xi(\pi)]\xi^+\,-\,e\xi\,({\cal V}_\mu
+{\cal A}_\mu)\,\xi^+\,\equiv\,{\bar \alpha}_{\mu\,\|} \,+\,{\bar
\alpha}_{\mu\,\perp}\,.\label{ALB}
\ee
Here ${\cal V}_\mu$ and ${\cal A}_\mu$ are the external vector and
axial vector fields, and
\bea
{\bar \alpha}_{\mu\,\|}\,&=&\,(2trS^a{\bar \alpha}_{\mu})\,S^a\,,\quad
{\bar \alpha}_{\mu\,\perp}\,=\,(2trX^a{\bar
\alpha}_{\mu})\,X^a\,,\quad
\xi(\pi)\,=\,exp(-i\pi(x)/f_\pi)\,, \nonumber \\
\pi(x)\,&=&\,\sum_a\,\pi^a(x)X^a\,, \\
 X^a\,&=&\,\frac{\tau^a}{2}\g5\,,\quad S^a\,=\,\frac{\tau^a}{2}\,. \label{xi}
\eea
\par
The current of our interest is presented in
Fig.\,\ref{figg1}. In order to derive the contribution of this
current to the space component of the WAEC, we need to extract
from the Lagrangian (\ref{LPNNL}) the
lowest order vertices
\be
\Delta {\cal L}_{\pi N}\,=\,-ig_A\,{\bar N}\,\gm\g5\,\frac{\vec \tau}{2}\,
N\cdot\vec{\cal A}_\mu\,-\,i\frac{g_A}{2f_\pi}\,{\bar N}\,\gm\g5\,\vec \tau\,
N\cdot\pa_\mu\vec{\pi}\,.                \label{DLPNNL}
\ee
%
%
\begin{figure}
\centerline{ \epsfig{file=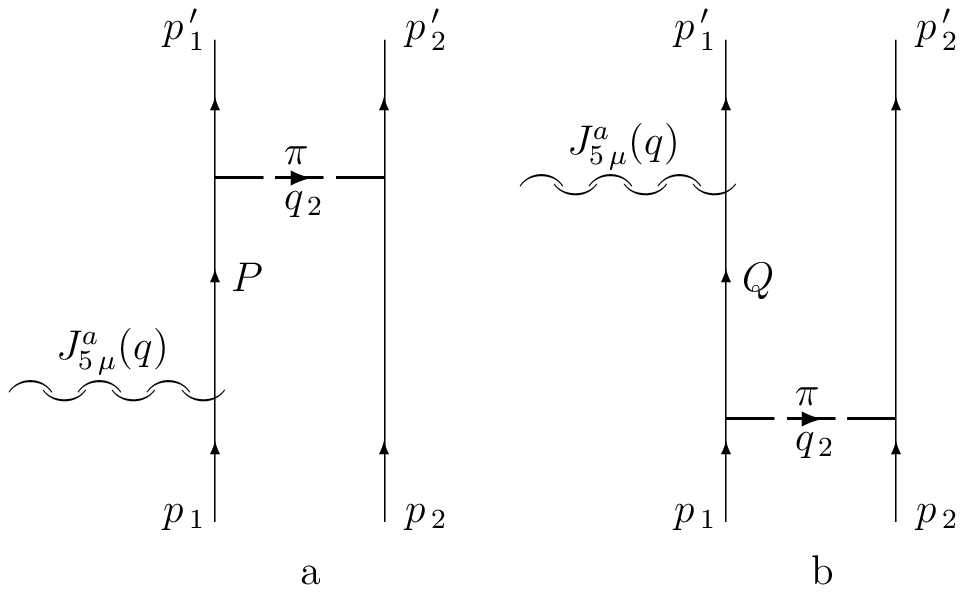}}
\vskip 0.4cm
\caption{ The weak axial nucleon Born term of the pion range. }
\label{figg1}
\end{figure}
%
%
\par
The Feynman amplitude reads
\bea
J^a_{\,5\mu}(pv)&\,=&\,-\bar{u}(p'_1)\left[\, \hat {\cal O}^\pi_{1}(-q_2)
S_F(P)\,{\hat J}_{\,5\mu}(1,q)\frac{1}{2}(a^+-a^-)
+{\hat J}_{\,5\mu}(1,q)\right.                      \nonumber \\
& & \left. \times\,S_F(Q)\hat {\cal O}^\pi_{1}(-q_2)\frac{1}{2}(a^++a^-)
\right] u(p_1) \Delta^\pi_F(q^2_2)\bar{u}(p'_2)\hat {\cal O}^\pi_{2}(q_2)
u(p_2)\,+\,\ot\,,                                   \label{JbBf}
\eea
where
\be
\hat {\cal O}^\pi(q_2)\,=\,\frac{f_{\pi NN}}{m_\pi}\,{\not q_2}\g5\,,\quad
 a^\pm\,=\,\frac{1}{2}\,[\tau^a_1,\,\tau^n_1]_\pm\,\tau^n_2\,,   \label{APM}
\ee
and we consider only the contact part of the one--nucleon
current
\be
{\hat J}_{\,5\mu}(1,c)\,=\,-ig_A\,\gm\g5\,. \label{OBC}
\ee
\par
In calculating the contribution of the amplitude
$J^a_{\,5\mu}(pv)$ to the exchange currents, one splits the
nucleon propagator into the positive- and negative frequency parts
and  the non--relativistic reduction is made. As  already
discussed in Sect.\,\ref{intro}, the contribution to the space
component of the negative frequency part of the Feynman amplitude
is of the nominal order ${\cal O}(1/M^{5})$ and therefore,
negligibly small. In the extended S--matrix method\footnote{The
same procedure has recently been applied in the study of the
$e$-$d$ scattering in Ref.\,\cite{PWD}.} \cite{ATA,RI},  first the
positive frequency part of the amplitude $J^a_{\,5\mu}(pv)$ is
written as
\bea
J^{a(+)}_{\,5\mu}(pv)\,&=&\,\frac{f_{\pi NN}}{m_\pi}\bar{u}(p'_1)
\left[\,\left(\qb\cdot{\vec\gamma}+iq_{20}\gamma_4\right)\g5
 \frac{1}{P_0-E(\vec P)} u(P){\bar u}(P)\,{\hat J}_{\,5\mu}(1,q)\,
 \frac{1}{2}(a^+ -a^-)\right.                              \nonumber \\
& & \left. \,+ {\hat J}_{\,5\mu}(1,q) \frac{1}{Q_0-E(\vec Q)}u(Q){\bar u}(Q)\,
 \left(\qb\cdot{\vec\gamma}+iq_{20}\gamma_4\right)\g5\,
 \frac{1}{2}(a^+ +a^-) \right]u(p_1)                        \nonumber \\
&&\,\times\,\Delta^\pi_F(q_2)\bar{u}(p'_2)\hat {\cal O}^\pi_{2}(q_2)
  u(p_2)+ \ot\,,                                            \label{JbBfP}
\eea
\par
It holds for the graph Fig.\,\ref{figg1}a
\be
q_{20}\,=\,P_0-p'_{10}\,=\,P_0-E(\vec P)\,+\,E(\vec P)-p'_{10}\,
\equiv\, P_0-E(\vec P)\,+\,q^{st}_{20}\,.   \label{Pq20st}
\ee
Similar equation holds for the graph Fig.\,\ref{figg1}b. Then
one obtains
\be
J^{a(+)}_{\,5\mu}(pv)\,=\,J^{a(+)}_{\,5\mu}(ps)\,+\, \Delta
J^{a}_{\,5\mu}(pv)\,.                        \label{JbBfPSDPV}
\ee
Here $J^{a(+)}_{\,5\mu}(ps)$ is the positive frequency part of the
nucleon Born term obtained using the static PS $\pi NN$ coupling.
It is the current containing a contribution from the potential,
since it coincides with the first Born iteration of the
Lippmann--Schwinger equation, if the static one--pion exchange
potential is used. In order to avoid double counting, the
contribution from such a graph is not included in the exchange
current since it is reducible.
\par
The current $\Delta J^{a}_{\,5\mu}(pv)$ arises from the contact
interaction
\bea
\Delta J^{a}_{\,5\mu}(pv)\,&=&\,i\frac{f_{\pi NN}}{m_\pi}\,
\bar u(p'_1)\left[\, \gamma_4\g5\,u(P)\bar u(P)\,{\hat J}_{\,5\mu}(1,q)
\frac{1}{2}(a^+ -a^-) \right.                           \nonumber \\
& &\left.\,-\, {\hat J}_{\,5\mu}(1,q)\,u(Q)\bar u(Q)\gamma_4\g5\,
\frac{1}{2}(a^+ +a^-)\,\right] u(p_1)                   \nonumber \\
&&\,\times\Delta^\pi_F(q_2)\bar{u}(p'_2)\hat {\cal O}^\pi_{2}(ps)
  u(p_2) + \ot\,,                                       \label{DJbBf}
\eea
where $\hat {\cal O}^\pi_{2}(ps)=ig_{\pi NN}\g5$.
The non--relativistic reduction of
the space component of the current (\ref{DJbBf}) yields
\bea
\Delta
{\vec j}^a_{\,5}(pv)\,&=&\,g_A\frac{g^2_{\pi NN}}{(2M)^3}\,
\left[\,\left(\vec q \,+\,i\sa\times\Pa\right){\,\tau^{\,a}_2}
+\,\left(i\Pa \,-\,\sa\times\vq\right)
 {\,\,(\vec \tau_1 \times \vec \tau_2)^a}\,\right]
\,\Delta^\pi_F(\qb^{\,\,2})(\sb\cdot\qb)                    \nonumber \\
&+& ~\ot\,.                                                 \label{DvJbBf}
\eea
where $\Pa={\vec p}_1+{\vec p}_1^{\,\,\prime}$. This current
coincides with the potential term derived earlier \cite{AHHST}
from the hard pion Lagrangian with the PV $\pi NN$ coupling
\cite{IT,ISTPR} and it contributes to the space component of the
WAEC in the same leading order in $1/M$ as other pion exchange
currents.
\par
The well known Foldy--Dyson unitary transformation of the nucleon
field \cite{F,Dy} can be used  in the Lagrangian (\ref{LPNNL}) to
obtain the PS $\pi NN$ coupling
\be
N\,=\,exp[-i\,\frac{g_A}{2f_\pi}\,\gamma_5\,(\vec \tau \cdot \vec \pi)]\,N'\,.
\label{FDT}
\ee
In this case, together with the nucleon Born term
$J^a_{\,5\mu}(ps)$ a contact amplitude $J^a_{\,5\mu}(PCAC)$,
called the PCAC constraint term, appears. For these amplitudes,
the following equation holds
\be
J^a_{\,5\mu}(pv)\,=\,J^a_{\,5\mu}(ps)\,+\,J^a_{\,5\mu}(PCAC)\,. \label{RIT}
\ee
It is clear that the resulting amplitude does not depend on the
nature of the $\pi NN$ coupling. This is due to the validity of
the powerful representation independence (equivalence) theorem
\cite{RH,DGH,FS}.
\par
In order to extract the nuclear WAEC from the relativistic
amplitudes in this case, the reducible part of the nucleon Born
amplitude $J^a_{\,5\mu}(ps)$ is isolated. This is the positive
frequency part $J^{a\,(+)}_{\,5\mu}(ps)$. Then from
Eqs.\,(\ref{JbBfPSDPV}) and (\ref{RIT}) we get
\be
\Delta J^{a}_{\,5\mu}(pv)\,=\,J^{a\,(-)}_{\,5\mu}(ps)\,+\, J^a_{\,5\mu}(PCAC)
      \,,                        \label{PVPS}
\ee
where $J^{a\,(-)}_{\,5\mu}(ps)$ is the negative frequency part of the
nucleon Born term obtained with the PS $\pi NN$ coupling.
Explicitly, one has for the  space component of the nuclear current,
given by the right hand side of Eq.\,(\ref{PVPS})
\be {\vec j}^a_{\,5}(ps)\,=\,g_A\frac{g^2_{\pi NN}}{(2M)^3}\,
\left[\,\left(\vec q \,+\,i\sa\times\Pa\right){\,\tau^{\,a}_2}
-\,(\sa\times\qb) {\,\,(\vec \tau_1 \times \vec
\tau_2)^a}\,\right] \,\Delta^\pi_F(\qb^{\,\,2})(\sb\cdot\qb)
\,+\,\ot\,,                                       \label{pspt} \ee
and
 \be {\vec j}^a_{\,5}(PCAC)\,=\,g_A\frac{g^2_{\pi
NN}}{(2M)^3}\, \left[\,i\Pa\,-\,(\sa\times\qa)\,\right] {\,\,(\vec
\tau_1 \times \vec \tau_2)^a}
\,\Delta^\pi_F(\qb^{\,\,2})(\sb\cdot\qb) \,+\,\ot\,.
\label{pcacct} \ee
\par
So in the chiral model with the PS $\pi NN$ coupling, the
potential current is obtained as the sum of  the negative
frequency part of the nucleon Born term and the PCAC constraint
term. This leads to the equality given by Eq.\,(\ref{DvJbBf}).
\par
The derivation of the WAEC from the hard pion Lagrangian with the
PS $\pi NN$ coupling was carried out earlier\,\cite{ISTPR,TK1}
with the following consequences:\\
(i) In a chiral invariant model with the PS $\pi NN$ coupling,
additional potential term arises, that makes the resulting current
equivalent to the current of the chiral model with the PV $\pi NN$
coupling. It follows the necessity of constructing the WAEC within
the chiral models and not simply in terms of $\pi NN$
couplings.\\
(ii) In order to avoid  double counting, the reducible part of the
potential current should be removed, since it is taken into
account already at the level of the impulse approximation
calculations. This procedure depends on the nuclear equation of
motion used for the description of nuclear states. Here the
calculation is carried out for the Schr\"odinger equation and
static one--pion exchange potentials. In our opinion, the problem
of double counting was omitted in \,\cite{Sci,TR,PMR}. Since the
potential term is absent in \,\cite{PMR}, those currents should be
used in conjunction with the Bethe-Salpeter equation, because, as
discussed in  ~\cite{KT1,D}, the WAEC does not contain the
contribution from the nucleon Born term in this case. On the other
hand, these currents \cite{PMR} are used at present in nuclear
physics calculations with wave functions derived with the
Schr\"odinger equation \cite{ASP,APKM,ASPFK}.
\par
Let us now discuss the continuity equation (\ref{PCAC}) for our current.
It can be shown that the nucleon Born term due to the contact part
of the one--body current (\ref{OBC}) of our model satisfies the
continuity equation
\be
q_\mu J^a_{5\mu,\pi}(B,c)\,=\,if_\pi\,M^a_\pi(B)\,, \label{NBT}
\ee
where $M^a_\pi(B)$ is the pion Born absorption amplitude given by
the graph of Fig.\,\ref{figg1} with the pion line instead of the
weak interaction wavy line inserted. Then the related nuclear
continuity equation for the nuclear current reads
\be
q_\mu j^a_{5\mu,\,\pi}(B,c)\,=\,i f_\pi {m}^a_\pi(2)\,
+\,\left(\left[V_\pi,\rho^a_5(1,c)\right]\,+\,\ot \right)\,,    \label{CPCAC}
\ee
Here the space part of the current
$j^a_{5\mu,\,\pi}(B,c)$ is given by Eq.\,(\ref{DvJbBf}) with the
divergence
\be \vq \cdot \Delta {\vec
j}^a_{\,5}(pv)\,=\,\frac{g_A^3}{8 M f_\pi^2}\, \left\{\,\left[\vec
q^{\,\,2} \,+\,i(\vq\cdot\sa\times\Pa)\right]{\,\tau^{\,a}_2}
+\,i(\vq\cdot\Pa) {\,\,(\vec \tau_1 \times \vec
\tau_2)^a}\,\right\}\,\Delta^\pi_F(\qb^{\,\,2})(\sb\cdot\qb)
\,+\,\ot\,, \label{dDvJbBf}
\ee
while it holds for the time
component that
\be
q_0\Delta j^{a}_{\,50}(pv)\,\approx\,{\cal O}(1/M^5)\,,   \label{dtc}
\ee
The pion absorption amplitude is obtained by the same method used
above for the derivation of the current $\Delta {\vec
j}_{\,5}(pv)$. Besides the contribution $m^a_\pi(2,ver)$ from the
energy dependence of the $\pi NN$ vertex of the internal pion, the
contribution $m^a_\pi(2,ext)$ from the energy dependence of the
$\pi NN$ vertex of the external pion arises with the result
\bea
if_\pi m^a_\pi(2,ver)&\,=\,&\vq \cdot \Delta {\vec j}^a_{\,5}(pv)\,,
                                                          \label{MAVER}\\
if_\pi m^a_\pi(2,ext)&\,=\,&\frac{g_A^3}{8 M f_\pi^2}\,
   \left\{\,\left[\vec q^{\,\,2}_2
   \,-\,i(\qb\cdot\sa\times\Pa)\right]{\,\tau^{\,a}_2}
   +\,i(\qb\cdot\Pa) {\,\,(\vec \tau_1 \times \vec \tau_2)^a}\,\right\}
   ~\Delta^\pi_F(\qb^{\,\,2})(\sb\cdot\qb)     \nonumber \\
  &+& ~\ot\,.  \label{MAEXT}
\eea
It is straightforward to obtain that the commutator of the static
one--pion exchange potential and the one--nucleon axial charge
density
\be
\rho^a_5(1,c)_i\,=\,\frac{g_A}{2M}({\vec \sigma}_i\cdot{\vec P}_i)
\frac{\tau^a}{2}\,,                        \label{ONCD}
\ee
is given by
\be
\left(\left[V_\pi,\rho^a_5(1,c)\right]\,+\,\ot \right)\,=\,-if_\pi
 m^a_\pi(2,ext)\,.                          \label{CVR}
\ee
The continuity equation (\ref{CPCAC})  which is in the leading
order in $1/M$ of the form
\be
\vq \cdot \Delta {\vec j}^a_{\,5}(pv)\,=\,i f_\pi
\left[m^a_\pi(2,ver)\,+\, m^a_\pi(2,ext)\right]\,
+\,\left(\left[V_\pi,\rho^a_5(1,c)\right]\,+\,\ot \right)\,,
\label{CPCACf}
\ee
is satisfied exactly. This is established from
Eqs.\,(\ref{dDvJbBf})-(\ref{CVR}). The contact term $\Delta
j^a_{\,5\mu}(pv)$ is related to the part of the continuity
equation, containing the potential and can be called as the true
potential current.
\par
Besides the nucleon Born term, our model Lagrangian contains a
${\cal A}\pi NN$ vertex
\be
\Delta{\cal L}_{{\cal A}\pi NN}\,=\,-\frac{i}{2f_\pi}{\bar N}\,\gm\,\vec \tau\,
 N \cdot \left(\vec\pi\times\vec{\cal A}_\mu\right)\,, \label{CAPINN}
\ee
providing another contact current that is a part of the full
contact term
\bea
j^a_{\,5\mu}(c)\,&=&\,\frac{i}{2f_\pi}\vamn\,\bar u(p'_1)\,
\left(\gm - \frac{\kappa^V}{2M}\sigma_{\mu\nu}q_\nu\right)\tau^m\,u(p_1)
\,\Delta^\pi_F(q^2)\,\bar{u}(p'_2)\hat {\cal O}^\pi_{2}(q_2)\tau^n u(p_2)
                                                 \nonumber \\
&+&\,\ot\,.                                      \label{NCT}
\eea
This current is required by the current algebra prediction for the
weak pion production  and it corresponds to the well known
$\rho$--$\pi$ current. It looks like a potential one, but it is
not connected to the potential and it satisfies the PCAC equation
\bea
q_\mu j^a_{\,5\mu}(c)\,&=&\,\frac{i}{2f_\pi}\vamn\,\bar u(p'_1)\,
\not q_2\tau^m\,u(p_1) \,\Delta^\pi_F(q^2)\,\bar{u}(p'_2)
\hat {\cal O}^\pi_{2}(q_2)\tau^n u(p_2)              \nonumber \\
\,&+&\,\ot\,\equiv\,if_\pi\,m^a_\pi(c)\,.            \label{dNCT}
\eea
The amplitude $m^a_\pi(c)$ is generated from the $NN\pi\pi$ term
$\Delta{\cal L}_{NN\pi\pi}=(i/4f^2_\pi){\bar N}\gm\vec\tau N\cdot
(\pa_\mu\vec \pi\times\vec\pi)$.
\par
Let us note that the contact term $j^a_{\,5\mu}(c)$,
Eq.\,(\ref{NCT}), is present in the HBChPT currents \cite{PKMR}
also.
\par
In the next section, we show  that the same potential current
(\ref{DvJbBf}) can be derived also within the HBChPT scheme.

\subsection{The weak axial pion pair term within the HBChPT formalism}
\label{secII.2}

We first derive the positive frequency nucleon Born term for the
weak pion production amplitude on  the nucleon in the leading
order. To this end, we start from the lowest order HBChPT
Lagrangian \cite{HHK,PMR,S}
\be {\cal L}^{(1)}_{\pi N}\,=\,-{\bar {\cal N}}_{v}\left[\,iv\cdot D\,
 +\,g_A\, S_v\cdot u\,\right]{\cal N}_v\,,  \label{LOLHBChPT}
\ee
where ${\cal N}_v$ is the velocity
dependent light component of the nucleon field $\Psi$, introduced
in the HBChPT and it is defined as
\be
{\cal N}_v\,\equiv\,e^{-iMv\cdot x}P_{v+}\,\Psi\,. \label{NVL}
\ee
Here
the four-velocity $v_\mu$ has the properties $v^2=-1$ and $v^0\ge
1$ and the projection operator $P_{v+}$ is defined as
\be
P_{v+}\,=\,\frac{1-i\not v}{2}\,.  \label{PVP}
\ee
For a choice
$v_\mu={p_\mu}/{M}$ we have
\be
P_{v+}\,=\,\frac{M-i\not p}{2M}\,.     \label{PVPs}
\ee
Taking into account  only the weak axial external interaction,
$a_\mu={\cal A}^a_\mu\,\tau^a/2$, we obtain in the leading order
\be
g_A S_v\cdot u\,\approx\,g_A \tau^a S_v\cdot {\cal A}^a_\mu\,
 -\,\frac{g_A}{f_\pi}\,S_{v,\mu}(\vec \tau \cdot \pa_\mu\vec\pi)\,.
 \label{APP}
\ee
Then the amplitude, corresponding to Fig.\,\ref{figg1}a reads
\be
M^v_c\,=\,-i\frac{2g_A^2}{f_\pi}N'N\tau^b\frac{\tau^a}{2}\,\bar u'_v\,
\left(S_v\cdot q_2\right)\frac{P_{v+}}{v\cdot K}
\left(S_v\cdot {\cal A}^a\right)\, u_v\,.                  \label{MVC}
\ee
Here
$N'$ and $N$ are the normalization factors. We use in $M^v_c$ the
choice
\be
v_\mu=p_\mu/M\,,\quad\,\vec p = \vec P \,,\quad\,p_0 = E(\vec P)\,.
\label{VMUSP}
\ee
With this choice,
the decomposition of the four-vector $P_\mu$ is
\be P_\mu\,=\,p_\mu\,+\,K_\mu\,.  \label{PMU} \ee so that the
scalar product ~$v\cdot K$~ in Eq.(\ref{MVC}) is given by
\be
v\cdot K\,=\,-v_0\,(P_0-E(\vec P))\,.  \label{vK}
\ee
Then we can write
\be
{\bar u}'_v\,\left(S_v\cdot q_2\right)\,=\,-\fot{\bar u}'_v\,\g5
({i\not q}_2\,+\,q_2\cdot v)\,=\,{\bar u}'_v\,
\left[\, S_v\cdot q^{st}_2\,-\fot\g5\,(P_0-E(\vec P))(\gamma_4+v_0)\,\right]
\,. \label{MVC1}
\ee
Employing Eqs.\,(\ref{vK}) and (\ref{MVC1}) in
Eq.\,(\ref{MVC}), we obtain
\bea
M^v_c\,&=&\,-i\frac{2g_A^2}{f_\pi}N'N\tau^b\frac{\tau^a}{2}\,\bar
u'_v\, \left(S_v\cdot q^{st}_2\right)\frac{P_{v+}}{v\cdot
K}\left(S_v\cdot {\cal A}^a\right)\, u_v            \nonumber \\
&-& ~i\,\frac{2g_A^2}{f_\pi}N'N\tau^b\frac{\tau^a}{2}\frac{1}{2v_0}\,\bar
u'_v\ \g5\gamma_4\,P_{v+}\left(S_v\cdot {\cal A}^a\right)\, u_v\,.
                                                    \label{MVCF}
\eea
For the amplitude, corresponding to Fig.\,\ref{figg1}b, we have
\bea
M^v_d\,&=&\,-i\frac{2g_A^2}{f_\pi}N'N\frac{\tau^a}{2}\tau^b\,\bar u'_v\,
\left(S_v\cdot {\cal A}^a\right)\frac{P_{v+}}{v\cdot K}
\left(S_v\cdot q^{st}_2\right)\, u_v                 \nonumber \\
&+& ~i\,\frac{2g_A^2}{f_\pi}N'N\frac{\tau^a}{2}\tau^b\frac{1}{2v_0}\,
\bar u'_v\ \left(S_v\cdot {\cal A}^a\right)P_{v+}\g5\gamma_4\, u_v\,.
                                                      \label{MVDF}
\eea
In $M^v_d$, we use the choice
\be
v_\mu=p_\mu/M\,,\quad\,\vec p=\vec Q\,,\quad\,p_0=E(\vec Q)\,.
\label{VMUSQ}
\ee
With this choice, the decomposition of the
four-vector $Q_\mu$ is
\be
Q_\mu\,=\,p_\mu\,+\,K_\mu\,,      \label{QMU}
\ee
from which it follows that
\be
v\cdot K\,=\,-v_0\,(Q_0-E(\vec Q))\,.  \label{vKQ}
\ee
Summing up the
partial results (\ref{MVCF}) and (\ref{MVDF}) we obtain
\be
M^v_{c+d}\,=\,M^v_{c+d}(st)\,+\,\Delta M^v_{c+d}\,, \label{MVDPC}
\ee
where
\bea
M^v_{c+d}(st) ~&=&~ -i ~\frac{2g_A^2}{f_\pi}N'N ~{\bar u}'_v
  \Biggr[ ~\left(S_v\cdot q^{st}_2\right) \frac{P_{v+}(P)}{v\cdot K}
  \left(S_v\cdot {\cal A}^a\right)~ \tau^b ~\frac{\tau^a}{2} \nonumber \\
&+& ~\left(S_v\cdot {\cal A}^a\right) ~\frac{P_{v+}(Q)}{v\cdot K}
     \left(S_v\cdot q^{st}_2\right)~ \frac{\tau^a}{2}\tau^b ~\Biggr] u_v ~,
                                                              \label{MVDPCS}
\eea
and
\bea
\Delta M^v_{c+d} ~&=&~ i ~\frac{g^2_A}{f_\pi}~ \frac{1}{v_0}N'N~
 {\bar u}'_v \Biggr[ ~\gamma_4\g5\,P_{v+}(P)~ \left(S_v\cdot {\cal A}^a\right)
 \tau^b ~\frac{\tau^a}{2}                                    \nonumber \\
 &-& ~\left(S_v\cdot {\cal A}^a\right)~ P_{v+}(Q)~ \gamma_4\g5~
      \frac{\tau^a}{2} \tau^b~ \Biggr] u_v ~.                \label{DMVDPC}
\eea
In order to obtain the two--nucleon amplitude, one should attach
the propagator of the intermediate meson and the ${\pi NN}$ vertex
of the second nucleon. According to the generalized Weinberg's
counting rules \cite{PMR}, such an amplitude has $\nu=-1$, like
the contact amplitude $j^a_{\,5\mu}(c)$, Eq.\,(\ref{NCT}). The
amplitude, following from  $M^v_{c+d}(st)$ is an analogue of the
positive frequency part of the nucleon Born term
$J^{a(+)}_{\,5\mu}(ps)$, obtained with the PS $\pi NN$ coupling.
In our opinion, only this part belongs to the class of reducible
diagrams that are not included in the exchange currents, if the
currents are used in conjunction with the Schr\"odinger equation
and the static one--pion exchange potential. On the other hand,
one can obtain from the interaction $\Delta M^v_{c+d}$ a contact
amplitude $\Delta J^a_{c+d,\mu}$
\bea
\Delta J^a_{c+d,\,\mu}\,&=&\,-\frac{g^3_A}{f^2_\pi}\frac{1}{v_{10}}N'_{v_1}
N_{v_1}{\bar u}'_{v_1}\left[\,\gamma_4\g5\,P_{v_1+}(P)S_{v_1,\mu}\,
\fot(a^+-a^-) \,-\,S_{v_1,\mu} P_{v_1+}(Q)\,\gamma_4\g5\,
\fot(a^++a^-)\right]u_{v_1}                              \nonumber \\
& &\,\times\,\Delta^\pi_F(q^2_2)\,N'_{v_2}N_{v_2}{\bar u}'_{v_2}
\,\left(S_{v_2}\cdot q_2\right)\,u_{v_2}\,,              \label{Dja}
\eea
that
is of the same form as $\Delta J^{a}_{\,5\mu}(pv)$ of
Eq.\,(\ref{DJbBf}). Making the non--relativistic reduction, one
obtains the contact term that is identical with the current
$\Delta j^{a}_{\,5\mu}(pv)$ of Eq.\,(\ref{DvJbBf}).

\subsection{Comparison of the WAEC} \label{secII.3}

Let us now compare the space component of the WAEC of the pion
range derived in the standard nuclear physics approach
\cite{IT,TK1,CT}, based on the chiral Lagrangians, with a similar
component in the HBChPT approach \cite{PMR,ASP} taken in the
leading order. The sum of the currents of the standard approach is
given by the contribution of the potential term as derived in
Sect.\,\ref{secII}, of the $\Delta(1232)$ isobar excitation and of
the $\rho$-$\pi$ current,
\bea
\vec j^a_{5,\,\pi}\,&=&\,\frac{g_A}{2M f^2_\pi}\,\left<\,
g^2_A\,\Big\{\Big(\frac{f_{\pi N \Delta}}
{f_{\pi NN}}\Big)^2\,\frac{2M}{9(M_\Delta - M)}\,\qb\,+\,\frac{1}{4}
[\vq+i(\sa\times \Pa)]\Big\}{\,\tau^{\,a}_2} \right.            \nonumber \\
& &\, \left. +\frac{1}{4}\,\left\{\Big[g^2_A\Big(\frac{f_{\pi N \Delta}}
{f_{\pi NN}}\Big)^2\,\frac{2M}{9(M_\Delta - M)}\,+\,(1+\kappa^V_\rho)\Big]
\,i(\sa\times \qb) \right.\right.                               \nonumber \\
& &\left.\left. \,+[g^2_A-(1+\kappa^V_\rho)]\,i(\sa\times \vq)
\,+\,(g^2_A-1)\Pa\right\}{\,i\,(\vec \tau_1 \times \vec \tau_2)^a}\,\right>
                                                                \nonumber \\
& &\,\times\,(\sb\cdot\qb)\,\Delta^\pi_F(\qb^{\,\,2})\,+\,\ot\,. \label{LRO}
\eea
The contribution from the $\Delta$ isobar excitation is specified by
the factor
$(\frac{f_{\pi N \Delta}}{f_{\pi NN}})^2/(M_\Delta -M)$,
where $M_\Delta$ is the mass of the $\Delta$ isobar and
$f_{\pi N \Delta}$ is the $\pi N \Delta$ coupling.
Other terms, containing $g^2_A$, are from the potential current.
In deriving Eq.\,(\ref{LRO}), we put the strong form factors
$F_{BNN}(\vec q^{\,\,2}_i)=1$,
$\Delta^\rho_F(\qa^{\,\,2})=1/m^2_\rho$ and we used the
Goldberger--Treiman and KSFR relations, $M|g_A|=g f_\pi$ and
$2 f^2_\pi g^2_\rho=m^2_\rho$, respectively.

The leading order HBChPT currents were compared \cite{ASP} with
the standard currents \cite{Sci} that contain the pion pair term
with the PS $\pi NN$ coupling. In comparing, this current was
omitted. The argument was that it corresponds to the PS $\pi NN$
coupling that is not chiral.
\par
Here we take for comparison the currents $\vec{A}^{\,\,a\,:\,\nu
3}_{12}(1\pi)$, [\cite{ASP}, (A5)], but with the potential current
(\ref{DvJbBf}) added. In our notation
\bea
\vec{A}^{\,\,a\,:\,\nu 3}_{12}(1\pi)\,&=&\,\frac{g_A}{2M f^2_\pi}\,\left<\,
\{ 2\hat{c}_3\,\vec q_2\,+\,\frac{g^2_A}{4}
[\vq+i(\sa\times \Pa)]\}{\,\tau^{\,a}_2} \right.                \nonumber \\
& &\left.\,+\frac{1}{4}\{(4 \hat{c}_4\,+\,1) \,i(\sa\times \qb) \,
+[g^2_A\,-\,1\,-\,c_6]\,i(\sa\times \vq)
\,+\,(g^2_A-1)\Pa\,\}{\,i\,(\vec \tau_1 \times \vec \tau_2)^a}\,\right>
                                                                \nonumber \\
& &\,\times\,(\sb\cdot\qb)\,\Delta^\pi_F(\qb^{\,\,2})\,+\,\ot\,. \label{EFTOLR}
\eea
\par
The currents $\vec j^a_{5,\,\pi}$ and $\vec{A}^{\,\,a\,:\,\nu
3}_{12}(1\pi)$ have an identical structure. This was achieved by
respecting  the chiral invariance and solving the double counting
problem in conjunction with the Schr\"odinger equation. In our
opinion, it is the current (\ref{EFTOLR}) that should be used in
the nuclear physics calculations with the nuclear wave functions
derived using the Schr\"odinger equation.
\par
Let us also note here that the weak axial exchange current
\cite{TR} can be used in conjunction with the equation of motion,
the first Born iteration of which coincides with the positive
frequency part of the nucleon Born term, constructed with the PV
$\pi NN$ coupling. In order to apply it in conjunction with the
Schr\"odinger equation, one should remove the reducible piece from
the positive frequency part of the nucleon Born term and add the
rest to the already derived exchange current \cite{TR}. The
resulting current will be of the order ${\cal O}(1/M^3)$. If the
pion exchange current \cite{TR} is constructed with the PS $\pi
NN$ coupling, according to the discussion after
Eq.\,(\ref{pcacct}), one should sum up the PCAC constraint term
and the negative frequency Born term (the pair term), both with
the potential imbedded. The resulting potential current will be
the same as in the PV $\pi NN$ coupling case.
\par
It follows also from the discussion after Eq.\,(\ref{pcacct}) that
one needs to add the PCAC constraint term to the pair term
\cite{Sci,ELM}, in order to obtain the chiral potential current.
\par
For the numerical estimate of the discussed effect, we compute the contribution
of the potential current to the cross section for the low energy electron
neutrino--deuteron inelastic scattering in the neutral current channel,
\be \nu_e\,+\,d\,\rightarrow\,\nu'_e\,+\,p\,+\,n\,.         \label{NUD}
\ee
This reaction is important for studying the solar neutrino
oscillations and it has been intensively studied both theoretically
\cite{MB2,ASPFK,NSGK,NSAPMGK,MRT,YHH}
and experimentally \cite{SNO1,SNO2}.
\par
The model axial current considered contains the one--nucleon
current and the WAEC (\ref{LRO}), to which we add also the
contribution from the $\Delta$ isobar excitation of the $\rho$
range. Referring to Sect.\,IV of Ref.\,\cite{MRT} for the details,
we present the results in table \ref{tab:1} and table \ref{tab:2}.
The nuclear wave functions are
generated by solving the Schr\"odinger equation with the Nijmegen
I potential \cite{SKTS} and the transition ${^3}S_1$-${^3}D_1
\rightarrow {^1}S_0$ was considered . The used weak interaction
constants are
$G_F=1.1803\times 10^{-5}$ GeV$^{-2}$, and $g_A=-1.267$. \\
\begin{table}
\caption{ Cumulative contributions to the cross section
$\sigma_{\nu d}$ ($\times 10^{-42}$ cm$^2$) from the weak axial
exchange currents  for various neutrino
energies are displayed. The cross section, calculated from  the sum of the
impulse approximation current (IA) and  of the
$\Delta$ isobar excitation of the $\pi$ and $\rho$ ranges is presented in
the row
labelled as +$\Delta(\pi+\rho)$. Other contributions correspond to
the $\rho$--$\pi$ current and to the pion potential term.
The cross section in the n-th row is given by the
contribution of all previous currents, the n-th current including. The
number in the bracket is the ratio of the n-th cross  section to
the cross section in the row above.}\label{tab:1}
\bigskip
\begin{tabular}{|l||c|c|c|c|c|}
\hline $E_\nu$ [MeV]& 5 & 10 & 15 & 20 & 101 \\\hline
IA       & 0.0938 (-)  &   1.076 (-) &  3.244  (-)  &  6.591 (-) &  147.1  (-)   \\
+$\Delta(\pi+\rho)$ & 0.0977 (1.041) & 1.126(1.046) & 3.405 (1.050) & 6.935(1.052)
                                                          & 158.5 (1.077) \\
+$\rho$-$\pi$ & 0.0986 (1.009) & 1.137 (1.010) & 3.443 (1.011)
                               & 7.016 (1.012) & 161.1 (1.016)        \\
+p($\pi$) & 0.0978 (0.992) & 1.127 (0.991) & 3.408 (0.990) & 6.940 (0.989)
                                           & 157.9 (0.980)            \\
\hline
\end{tabular}
\end{table}
It is seen from table \ref{tab:1} that the effects of the $\rho$-$\pi$ and
potential terms are $\sim$ 1 \% and they cancel each other to a
large extent. Since the total  effect from the space part of WAEC
is at the level of a few percent, it is important to correctly
identify all the components of the WAEC that satisfy the PCAC and
contribute sensibly.

In Ref.\,\cite{MB2}, the effective cross section for the
reaction (\ref{NUD}) is  presented in the form \be
\sigma_{EFT}(E_\nu)\,=\,a(E_\nu)\,+\,L_{1,\,A}\,b(E_\nu)\,.
\label{sigB} \ee The amplitudes $a(E_\nu)$ and $b(E_\nu)$ are
tabulated in \cite{MB2} from the neutrino
energy 3 MeV up to 20 MeV, with the step 1 MeV. The constant $L_{1,\,A}$
cannot be determined from reactions between elementary particles.
Here we extract $L_{1,\,A}$ from our cross sections. The used weak interaction
constants are now
$G_F=1.166\times 10^{-5}$ GeV$^{-2}$, and $g_A=-1.26$. The results are presented in table \ref{tab:2}.

\begin{table}
\caption{ The values of the constant  $L_{1,\,A}$ obtained by the
fit to the cross section of the reaction (\ref{NUD})
calculated using the NijmI potential.} \label{tab:2}
\begin{center}
\begin{tabular}{|l||c|c|c|c|}\hline 
                  & IA  & +$\Delta(\pi+\rho)$  & +$\rho$-$\pi$ & +p($\pi$)   \\\hline\hline
${L}_{1,\,A}$ & 0.8 & 4.1 & 4.9 &   4.2   \\\hline
\end{tabular}
\end{center}
\end{table}

It is seen from table \ref{tab:2} that the effects of the $\rho$-$\pi$ and
potential terms change the value of the constant  ${L}_{1,\,A}$ sensibly and
that again, they cancel each other to a large extent.

\section{Results and Conclusions} \label{secIII}

The question of the interplay of the chiral invariance restriction
and of the double counting problem in the construction of the weak
axial potential exchange currents of the pion range is discussed.
It is shown that in order to avoid the double counting problem,
one should study the structure of the first Born iteration of the
nuclear equation of motion and of the nucleon Born term. Only the
part of the nucleon Born term, that is not contained in the first
Born iteration contributes to the exchange currents. This current
was constructed in conjunction with the Schr\"odinger equation in
Sect.\,\ref{secII}. Then it is shown that the total potential
exchange current, with the pion pair term included, satisfies the
PCAC constraint (\ref{NCEt}). The construction is done in the
leading order both in the chiral perturbation theory and in the
HBChPT approach. The resulting potential term is the same in both
approaches and it coincides with the potential term derived
earlier from the hard pion Lagrangians. It is also shown that with
the correct potential term taken into account, the leading order
part of the space component of the long--range weak axial exchange
currents of the HBChPT approach is identical with such a component
obtained within the standard nuclear physics approach based on
chiral Lagrangians. The same is also true for the pion exchange
currents constructed in Refs.\,\cite{Sci,TR}.
\par
Numerically, the contribution of the potential term is at the same
level as the contribution from the well--known $\rho$--$\pi$
current and the two contributions tend to cancel each other at low
energies.
\par
Let us note that in Ref.\,\cite{JW}, the time component of the
electromagnetic exchange currents of the pion range was
constructed in conjunction with the
Blankenbecler--Sugar--Logunov--Tavkhelidze equation
\cite{BlSu,LT}, that is a 3-dimensional reduction of the
Bethe--Salpeter equation\footnote{See also Ref.\,\cite{CR}.}. It
was shown \cite{GA} that the resulting exchange charge density is
equivalent to that obtained by such standard methods as are the
transformation method \cite{GA} and the extended S--matrix method
\cite{ATA} and that it is independent of the form of the $\pi NN$
coupling. This result provides a strong argument that the WAEC of
the pion range constructed here, and more generally, the
one--boson WAEC constructed in Ref.\cite{MRT} from chiral
Lagrangians, can be used in standard nuclear physics calculations
also in conjunction with the corresponding Lippmann--Schwinger
equation, obtained \cite{Mac} by the above discussed reduction of
the Bethe--Salpeter equation, and using the Bonn potentials
\cite{Mac,OPT,CDB} for generating the nuclear wave functions.

\section*{Acknowledgments}

This work is supported in part by the grant GA CR 202/03/0210 and
by Ministero dell' Istruzione, dell' Universit\`a e della Ricerca
of Italy (PRIN 2003). We thank J. Smejkal for discussions
and F.C. Khanna for the critical reading of the manuscript.

%
%

\end{document}